\RequirePackage{fix-cm}
\documentclass[smallextended]{svjour3}       
\smartqed  
\usepackage{graphicx}
\begin{document}

\title{Constructions of Optimal and Near-Optimal Quasi-Complementary Sequence Sets from an Almost Difference Set \thanks{The work is supported by Shandong Province Natural Science Foundation of China(No.ZR2016FL01), The Fundamental Research Funds for the Central Universities(No.16CX02013A, No. 17CX02030A, No. 15CX02056A), Qingdao application research on special independent innovation plan project(No. 16-5-1-5-jch)}}



\author{Yu Li \and Tongjiang Yan \and Chuan Lv}


\institute{Yu Li \at
               College of Sciences,
China University of Petroleum,
Qingdao 266555,
Shandong, China\\
\at Fujian Provincial Key Laboratory of Network Security and Cryptology, Fujian Normal University Fuzhou 350007, China\\
\email{liyu$1992$@outlook.com}
\and
Tongjiang Yan\at
College of Sciences,
China University of Petroleum,
Qingdao 266555,
Shandong, China\\
\email{yantoji@163.com}
\and
Chuan Lv\at
College of Sciences,
China University of Petroleum,
Qingdao 266555,
Shandong, China\\
\email{lvchuan501@163.com}
}
\date{Received: date / Accepted: date}

\maketitle

\begin{abstract}
Compared with the perfect complementary sequence sets, quasi-complementary sequence sets (QCSSs) can support more users to work in multicarrier CDMA communications. A near-optimal periodic QCSS is constructed in this paper by using an optimal quaternary sequence set and an almost difference set. With the change of the values of parameters in the almost difference set, the near-optimal QCSS can become asymptotically optimal and the number of users supported by the subcarrier channels in CDMA system has an exponential growth.
\end{abstract}

\section{Introduction}
\label{section 1}
A two-dimensional matrix is called a perfect complementary sequence (PCS) if the summation of its auto-correlations of the row sequences is to zero for all non-zero time-shifts. In 1961, Golay first proposed a pair of sequences for the aperiodic PCS \cite{01}, then Tseng and Liu generalized it to more than two sequences, B\"omer and L\"uke proposed the periodic PCS and odd-periodic PCS, respectively \cite{02,04,03}. A perfect complementary sequence set (PCSS) is a set of mutually orthogonal complementary sequences, whose cross-correlations sum to zero \cite{06,05}. Nowadays the PCSS has sought an important application in multicarrier code-division multiple-access (MC-CDMA) communication \cite{07,08}. However the small set size of PCSS has a remarkable limitation for the number of users in CDMA communication. This is for each PCSS, the number of subcarrier channels, denoted by $M$, is an upper bound of the number of CDMA users that can be supported, denoted by $K$, \emph{i.e.}, $K\leq M$. This limitation can be overcome by allowing the periodic tolerance (or the maximum periodic correlation magnitude) to take low non-zero values, so the quasi-complementary sequence set (QCSS) is proposed in \cite{11,09,10}.

In \cite{12}, the author proposed two optimal and near-optimal periodic QCSSs by using the singer difference set and existing optimal quaternary sequence sets established by Sol, Boztas \emph{et al}, and Udayia \emph{et al} \cite{13,15,14}, which can support more CDMA users, \emph{i.e.}, $K\geq2M$. In this paper, we will propose a near-optimal QCSS by using existing optimal quaternary sequence set and an almost difference set derived from the singer difference set \cite{19,20,18}. Note that the parameters of the QCSS proposed by the singer difference set are extremely fixed \cite{12}. Once there are some slight changes for the parameters, the optimal/near-optimal properties will disappear. However, in this paper, with the reasonable changes of the values of parameters in the almost difference set, the near-optimal QCSS proposed by the almost difference set will become asymptotically optimal and the number of CDMA users that can be supported by proposed QCSS will have an exponential growth.

The rest of this paper is organized as follows. In Section 2, we introduce some necessary notations and well-known results. In Section 3, we present the framework for the construction of the optimal QCSS, following by the proposed constructions for the near-optimal periodic QCSS. By analysing numerical results of the correlation lower bound, the periodic asymptotically optimal QCSS is constructed. And we will summarize and give some remarks about our results in Section 4.

\section{Preliminaries}\label{section 2}

Given two complex-valued sequences $a=\{a_{t}\}$ and $b=\{b_{t}\}$ of length $N$, the periodic correlation function of $a$ and $b$ is defined as
\begin{equation}
R(a, b; \tau)=\sum_{t=0}^{N-1}a_{t}b_{t+\tau}^*,\ 0\leq \tau\leq N-1,
\end{equation}
where the addition $t+\tau$ is performed modulo $N$, $b_{t+\tau}^*$ denotes the Hermite transposition of $b_{t+\tau}$. If $a\neq b$, $R(a, b; \tau)$ is called the periodic cross-correlation function (PCCF), otherwise, it is called the periodic auto-correlation function (PACF). For simplicity, the PACF of $a$ can be sometimes written as $R(a;\tau)$.
\subsection*{\emph{2.1\ Periodic Quasi-Complementary Sequence Set.}}

Let a set $C=\{C_{0},C_{1},\cdots,C_{K-1}\}$ contain $K$ two-dimensional matrices, each of size $M\times N$, \emph{i.e.},
\begin{equation}
C_{k}=\left[
\begin{array}{c}
C_{0}^k \\
C_{1}^k \\
\vdots \\
C_{M-1}^k\\
\end{array}
\right]_{M\times N}, \mathrm{for}\ 0\leq k\leq K-1.
\end{equation}
The periodic correlation function of $C$, which is in the form of the correlation sum, is defined as
\begin{equation}
R(C_{k_{1}}, C_{k_{2}}; \tau)=\sum_{m=0}^{M-1}R(C_{m}^{k_{1}}, C_{m}^{k_{2}}; \tau),
\end{equation}
where $0\leq k_{1},k_{2}\leq K-1$.

\begin{definition}\label{definition 1}
\cite{12}The periodic auto-correlation tolerance $\delta_{a}$ and the periodic cross-correlation tolerance $\delta_{c}$ are defined as
\begin{eqnarray*}
\delta_{a}&=&max\{|R(C_{k}; \tau)|: 0< |\tau|\leq N-1\},\\
\delta_{c}&=&max\{|R(C_{k_{1}}, C_{k_{2}}; \tau)|: k_{1}\neq k_{2}, 0\leq |\tau|\leq N-1\},
\end{eqnarray*}
respectively. Moreover, the periodic tolerance (or the maximum periodic correlation magnitude) $\delta_{max}$ of $C$ is defined as
\begin{equation}\label{periodic correlation}
\delta_{max}=max\{\delta_{a}, \delta_{c}\}.
\end{equation}
Obviously, $\delta_{max} \geq 0$.
\end{definition}
\begin{definition}\label{definition 2}
If the periodic tolerance $\delta_{max}>0$ in Eq.(\ref{periodic correlation}), the set $C$ is called a $(K,M,N)$ periodic quasi-complementary sequence set, denoted by $(K,M,N)$-QCSS, where $K$ denotes the size of the set $C$, and $M$ and $N$ denote the number of rows and lines of each element matrix in the set $C$, respectively.
\end{definition}

\begin{lemma}\label{lower bound}
\cite{12}(Correlation lower bound for a periodic QCSS) For a polyphase periodic $(K,M,N)$-QCSS, we have
\begin{equation}\label{low_bound}
\delta_{max}^2\geq M^2N^2\frac{K/M-1}{KN-1}.
\end{equation}
\end{lemma}

\begin{remark}\label{remark 1}
If $M=1$, the element $C_{k}(0\leq k\leq K-1)$ of the set $C$ will reduce to a conventional sequence. For ease of presentation, in this case, the periodic tolerance $\delta_{max}$ is denoted by $\alpha_{max}$, then its lower bound is shown below
\begin{equation}
\alpha_{max}^2\geq N^2\frac{K-1}{KN-1}.
\end{equation}
\end{remark}

In a multicarrier CDMA system with $M$ subcarrier channels and $N$ chip slots, the QCSS with parameters $K,M$ and $N$ is able to support $K$ users to communicate. More specifically, each data signal of a specific user is spread by a complementary matrix in frequency and time domains, \emph{i.e.}, each row sequence of the element of the QCSS is sent out over one of the $M$ subcarrier channels and all the row sequences are simultaneously sent out over $N$ chip slots. For more detail, the reader is referred to \cite{07} and \cite{08}.

\subsection*{\emph{2.2\ Almost Difference Set (ADS)}}

Let $(A, +)$ be an Abelian group of order $P$. Let $D$ be a $M$-subset of $A$. The set $D$ is an $(P, M, \lambda, t)$ almost difference set of $A$, denoted by ADS, if the difference function $d_{D}(x)$ takes on $\lambda$ altogether $t$ times and $\lambda +1$ altogether $P-1-t$ times when $x$ ranges over all the nonzero elements of $A$; if $t=P-1$, the set $D$ is called a $(P, M, \lambda)$ difference set of $A$, denoted by DS, where the difference function $d_{D}(x)$ is defined by
\begin{eqnarray*}
d_{D}(x)=|(D+x)\cap D|.
\end{eqnarray*}
If there exits an $(P, M, \lambda, t)$ almost difference set, we have
\begin{eqnarray*}
M(M-1)=t\lambda +(P-1-t)(\lambda +1).
\end{eqnarray*}
\begin{lemma}\label{ADS}
\cite{18} Let $W$ be any $(f,\frac{f-1}{2},\frac{f-3}{4})$ or $(f,\frac{f+1}{2},\frac{f+1}{4})$ difference set of $\texttt{Z}_{f}$, where $f\equiv 3(\bmod 4)$. Define a subset of $\texttt{Z}_{4f}$ by
\begin{eqnarray*}
U&=&[(f+1)W\ mod\ 4f]\cup [(f+1)(W-\delta)^*+3f\ mod\ 4f]\cup \\
&&[(f+1)W^*+2l\ mod\ 4f]\cup[(f+1)(W-\delta)^*+3f\ mod\ 4f],
\end{eqnarray*}
where $W^*$ and $(W-\delta)^*$ denote the complements of $W$ and $W-\delta$ in $\texttt{Z}_{f}$, respectively. Thus $U$ is an $(4f, 2f-1, f-2, f-1)$ or $(4f, 2f+1, f, f-1)$ almost difference set of $\texttt{Z}_{4f}$.
\end{lemma}

\subsection*{\emph{2.3\ Optimal Quaternary Family $\mathcal{A}$}}

In this part, an optimal quaternary sequence set, Family $\mathcal{A}$, is given a brief review. In 1988, the optimal quaternary Family $\mathcal{A}$ was originally discovered by P.Sol\'e, which comprises of $2^n+1$ $\texttt{\textbf{Z}}_{4}$ sequences of period $2^n-1$ \cite{13}. Let $f(x)=x^n+a_{n-1}x^{n-1}+\cdots+a_1x+a_0$ be a primitive basic irreducible polynomial of degree $n$ and divide $x^{2^n-1}-1$ in $\texttt{\textbf{Z}}_{4}[x]$, whose modulo 2 projections are primitive irreducible polynomials in $\texttt{\textbf{Z}}_{2}[x]$. Consider the $n$th-order linear recurrence over $\texttt{\textbf{Z}}_{4}$ having characteristic polynomial $f(x)$ as follows
\begin{eqnarray}\label{A}
s(t)+a_{n-1}s(t-1)-a_{n-2}s(t-2)+ \cdots +a_{0}s(t-n)=0.
\end{eqnarray}
Let $S(f)$ denote the set of all sequences over $\texttt{\textbf{Z}}_{4}$ satisfying E.q(\ref{A}), and $S^+(f)$ is the set of all nonzero solutions. Family $\mathcal{A}$ then defined to be the family $S^+(f)$ provided $f(x)$ in $\texttt{\textbf{Z}}_{4}$ \cite{14}. For more information, the readers are referred to \cite{15,16,17}. We present the following correlation properties of Family $\mathcal{A}$.
\begin{lemma}\label{Family A}
\cite{12}The Family $\mathcal{A}=\{l_{0}, l_{1}, \cdots, l_{2^n} \}$ consists of $2^n+1$ sequences, each of which has a length of $2^n-1$. In particular, $l_{0}$ is a binary $m$-sequence which has non-trivial auto-correlation values of $-1$.
The periodic tolerance of the Family $\mathcal{A}$ is
\begin{equation}
\alpha_{max}=1+2^{n/2}.
\end{equation}
The subset $\mathcal{L}=\{l_{1}, l_{2}, \cdots, l_{2^n} \}$ of $\mathcal{A}$ has the following property
\begin{equation}
R(l_{k_{1}}, l_{k_{2}}; 0)=-1,
\end{equation}
where $1\leq k_{1}\neq k_{2}\leq 2^n$.
\end{lemma}

\section{Proposed Construction for Optimal and Near-Optimal QCSS}\label{section 3}

In this section, we firstly propose a framework of the QCSS based on a sequence set and an integer set by a linear mapping, meanwhile, the periodic tolerance and the tightness factor of the QCSS are considered. Then we construct the optimal and near-optimal QCSSs by applying an optimal quaternary sequence set, Family $\mathcal{A}$, and an almost difference set to this framework, and analyze their asymptotic properties.

\subsection*{\emph{3.1\ Framework for the Construction of QCSS}}

For any length-$N$ complex-valued sequence $a$, define the following mapping as the linear phase transform of $a$, \emph{i.e.},
\begin{eqnarray*}
\varphi(a,d)=(a_{0}\xi_{q}^{0d},a_{1}\xi_{q}^{1d},\cdots,a_{N-1}\xi_{q}^{(N-1)d})),
\end{eqnarray*}
where $d$ and $q$ are integers, and $\xi_{q}=exp\{2\pi\sqrt{-1} /q\}$.

\begin{lemma}\label{linear R}
\cite{12}For two length-$N$ complex-valued sequences $a$ and $b$, we have
\begin{equation}\label{lin_R}
R(\varphi(a,d), \varphi(b,d); \tau)=\xi_{q}^{-\tau d}R(a,b;\tau).
\end{equation}
\end{lemma}
Suppose there exists a sequence set $v=(v_{0}, v_{1}, \cdots, v_{K-1})$ with the periodic tolerance $\alpha_{max}$, which has $K$ length-$N$ sequences, and an integer set $D=\{d_{0}, d_{1}, \cdots, d_{M-1} \}$, then we can construct a QCSS
\begin{eqnarray}\label{C}
C=\{C_{0}, C_{1}, \cdots, C_{K-1} \},
\end{eqnarray}
where
\begin{eqnarray*}
C_{k}=\left[
\begin{array}{c}
\varphi_{q}(v_{k},d_{0}) \\
\varphi_{q}(v_{k},d_{1}) \\
\vdots \\
\varphi_{q}(v_{k},d_{M-1})\\
\end{array}
\right]_{M\times N}, \mathrm{for}\ 0\leq k\leq K-1.
\end{eqnarray*}
Based on Eq.(\ref{lin_R}), for $0\leq k_{1}, k_{2}\leq K-1$,
\begin{eqnarray}\label{R}
& &|R(C_{k_{1}}, C_{k_{2}}; \tau)| \nonumber \\
&=&|R(v_{k_{1}}, v_{k_{2}}; \tau) \sum_{m=0}^{M-1}\xi_{q}^{-\tau d_{m}}|.
\end{eqnarray}
Next, we continue the discussion into two cases: \\
$1)$ For $\tau \not\equiv 0(\bmod \ q)$, let
\begin{eqnarray}\label{R1}
&&max|R(C_{k_{1}}, C_{k_{2}}; \tau)| \nonumber \\
&=&\alpha_{max}|\sum_{m=0}^{M-1}\xi_{q}^{-\tau d_{m}}| \nonumber \\
&=&R_{1}.
\end{eqnarray}
$2)$ For non-trivial $\tau \equiv 0(\bmod \ q)$, let
\begin{eqnarray}\label{R2}
&&max|R(C_{k_{1}}, C_{k_{2}}; \tau)|  \nonumber \\
&=&M|\sum_{m=0}^{M-1}\xi_{q}^{-\tau d_{m}}|  \nonumber \\
&=&R_{2}.
\end{eqnarray}

To analyze the tightness of Eq.(\ref{low_bound}), define the tightness factor $\rho$ as
\begin{eqnarray}\label{tightness}
\rho=\frac{\delta_{max}}{MN\sqrt{\frac{K/M-1}{KN-1}}},
\end{eqnarray}
which is a measure of closeness between the periodic tolerance and the periodic correlation lower bound in Lemma \ref{lower bound}.

\begin{remark}\label{remark 3}
In general, $\rho\geq 1$. If $\rho=1$, the proposed $C$ is said to be optimal; if $\rho$ is a small number, \emph{i.e.}, $1< \rho\leq 2$, the proposed $C$ is said to be near-optimal.
\end{remark}

\begin{lemma}\label{upper-bound}
Let set $D=\{d_0, d_1, \cdots, d_{M-1}\}$ be an $(q, M, \lambda, t)$ almost difference set in $\texttt{Z}_{q}$. Then
\begin{eqnarray*}
|\sum_{m=0}^{M-1}\xi_{q}^{-\tau d_{m}}|<\sqrt{M+q-\lambda -1},
\end{eqnarray*}
where $\tau$ is an integer.
\end{lemma}
$\mathbf{Proof.}$ Define a FFT as follows,
\begin{eqnarray}
f(\tau)\ \colon=\frac{1}{M}|\sum_{m=0}^{M-1}\xi_{q}^{\tau d_{m}}|,
\end{eqnarray}
and
\begin{eqnarray*}
x[\tau]\ &\colon=&M^2(f_{\tau}^2 -\frac{q-M}{(q-1)M})=\frac{M(M-1)}{q-1}+\sum_{l=1}^{q-1}a_{l}\xi_{q}^{\tau l}, \\
a_{0}\ &\colon=&\frac{M(M-1)}{q-1}, \\
\end{eqnarray*}
where the $a_{l}$ denotes the number of occurrences of $l$, $l=1, 2, \cdots, q-1$, and we have $x[\tau]=\sum\limits_{l=0}^{q-1}a_{l}\xi_{q}^{\tau l}$.
According to the property of FFT,
\begin{eqnarray*}
\sum_{l=0}^{q-1}\xi_{q}^{\tau l}&=&\sum_{l=0}^{q-1}e^{2\pi \tau l/q} \\
&=&\sum_{l=0}^{q-1}\cos \frac{2\pi \tau l}{q}+i\sum_{l=0}^{q-1}\sin \frac{2\pi \tau l}{q} \\
&=&0.
\end{eqnarray*}
This is equivalent to
\begin{eqnarray*}
\cos0+i\sin0+\sum_{l=1}^{q-1}\cos \frac{2\pi \cdot \tau l}{q}+i\sum_{l=1}^{q-1}\sin \frac{2\pi \tau l}{q}=0.
\end{eqnarray*}
Thus,
\begin{eqnarray*}
\sum_{l=1}^{q-1}\cos \frac{2\pi \tau l}{q}&=&-1, \\
\sum_{l=1}^{q-1}\sin \frac{2\pi \tau l}{q}&=&0.
\end{eqnarray*}
With the definition of ADS,
\begin{eqnarray*}
a_{l_{h}}&=&\lambda,\ \ h=1, 2, \cdots, t \\
a_{l_{j}}&=&\lambda+1,\ \ j=1, 2, \cdots, q-1-t.
\end{eqnarray*}
Hence,
\begin{eqnarray*}
x[\tau]&=&a_{0}+\lambda \sum_{h=1}^{t}\cos \frac{2\pi \tau l_{h}}{q} + i\lambda \sum_{h=1}^{t}\sin \frac{2\pi \tau l_{h}}{q} \\
&&+(\lambda +1)\sum_{j=1}^{q-1-t}\cos \frac{2\pi \tau l_{j}}{q} + i(\lambda +1)\sum_{j=1}^{q-1-t}\sin \frac{2\pi \tau l_{j}}{q} \\
&=&a_{0}+\lambda \sum_{l=1}^{q-1}\cos \frac{2\pi \tau l}{q}  +\sum_{j=1}^{q-1-t}\cos \frac{2\pi \tau l_{j}}{q}  +i\sum_{j=1}^{q-1-t}\sin \frac{2\pi \tau l_{j}}{q} \\
&=&a_{0}-\lambda  +\sum_{j=1}^{q-1-t}\cos \frac{2\pi \tau l_{j}}{q}.
\end{eqnarray*}
Because of $|\cos \theta| \leq 1,\forall \theta \in R$, then
\begin{eqnarray}\label{x_tau}
a_{0}-\lambda-(q-1-t)\leq x[\tau]\leq a_{0}-\lambda+(q-1-t).
\end{eqnarray}
In general, the bounds in Eq.(\ref{x_tau}) is not obtained,\emph{ i.e.},
\begin{eqnarray*}
a_{0}-\lambda-(q-1-t)< x[\tau]< a_{0}-\lambda+(q-1-t).
\end{eqnarray*}
Let $\Delta =Mf_{\tau}$, then
\begin{eqnarray*}
x[\tau]=\Delta ^2-\frac{M(q-M)}{q-1},
\end{eqnarray*}
thus,
\begin{eqnarray*}
\Delta ^2=x[\tau]+\frac{M(q-M)}{q-1}
\end{eqnarray*}
and
\begin{eqnarray}\label{Delta}
\Delta_{max}&\leq&\sqrt{a_{0}-\lambda+q-1-t+\frac{M(q-M)}{q-1}} \nonumber\\
&=&\sqrt{q+M-\lambda -t-1}.
\end{eqnarray}
Obviously, the bound in Eq.(\ref{Delta}) is not obtained.

\subsection*{\emph{3.2\ Proposed Near-Optimal QCSS $C^0$ from the Almost Difference Set and the Family $\mathcal{A}$}}

\begin{theorem}\label{n-p-QCSS}
For an $(q, M, \lambda, t)$-ADS and the subset $\mathcal{L}=\{l_{1}, l_{2}, \cdots, l_{2^n}\}$ of Family $\mathcal{A}$, a near-optimal $(K, M, N)$-QCSS $C^0$ can be constructed, where $K=2^n,\ M=2^{n-1}-3,\ N=2^n-1,\ n>2$.
\end{theorem}
$\mathbf{Proof.}$ Considering $\mathcal{L}=\{l_{1}, l_{2}, \cdots, l_{2^n}\}$, which is the subset of Family $\mathcal{A}$, based on Lemma \ref{Family A}, we have
\begin{eqnarray}
N=2^n-1,\ \ K=2^n,\ \ \alpha_{max}=1+2^{n/2}.
\end{eqnarray}

For $f=2^{n-2}-1$, based on Lemma \ref{ADS}, we have
\begin{eqnarray}\label{ASD q}
q=2^n-4,\ M=2^{n-1}-3,\ n\geq 3,
\end{eqnarray}
and there exists an $(2^n-4,\ 2^{n-1}-3,\ 2^{n-2}-3,\ 2^{n-2}-2)$-ADS.\\
1)For $\tau \not\equiv 0 \ (\bmod \ 2^n-4)$, by Eq.$(\ref{R1})$,
\begin{eqnarray}
R_{1}&=&max|R(C_{k_{1}}, C_{k_{2}}; \tau)| \nonumber\\
&=&\alpha_{max} max|\sum_{m=0}^{M-1}\xi_{q}^{-\tau d_{m}}| \nonumber\\
&=&\alpha_{max}\Delta_{max} \nonumber\\
&\leq&(1+2^{n/2})\sqrt{q+1} \nonumber\\
&=&(1+2^{n/2})\sqrt{2^{n}-3}
\end{eqnarray}
2)For non-trivial $\tau \equiv 0 \ (\bmod \ 2^n-4)$, by Eq.$(\ref{R2})$,
\begin{eqnarray}
R_{2}&=&max|R(C_{k_{1}}, C_{k_{2}}; 0)| \nonumber\\
&=&M \nonumber\\
&=&2^{n-1}-3.
\end{eqnarray}
Note that $R_{1}>R_{2}$. Substituting the values of $R_{1},K,M,N$ into Eq.(\ref{tightness}), the value of $\rho$ can be gained as follows,
\begin{equation}\label{rho_odd_even}
\rho \leq \left\{
\begin{array}{cc}
\sqrt{2},& \ \ \emph{if\ n\ is\ odd},\\
2,& \ \ \ \emph{if\ n\ is\ even}.
\end{array}
\right.
\end{equation}
According to Eq.(\ref{Delta}), the bounds are not obtained in Eq.(\ref{rho_odd_even}), thus
\begin{equation}\label{rho}
1< \rho <2.
\end{equation}
By Eq.(\ref{rho}), and recalling Remark \ref{remark 3}, we assert that $C^0$ is asymptotically near-optimal and $K\geq 2M$.

\begin{remark}\label{DS}
In \cite{12}, with the increase of $x$ about $q=2^{n-x}$ for all possible $n$ of the $(q,\frac{q-1}{2},\frac{q-3}{4})$ singer difference set in proposed $C^1$, the value of $\rho$ has a gradual increasing trend, \emph{i.e.}, the optimal property of the proposed QCSS would gradually disappear. The numerical analysis is shown in Table 1.
\end{remark}
\begin{remark}\label{result}
In this paper, with the increase of $x$ about $f=2^{n-x}$ for all possible $n$ of $(4f, 2f-1, f-2, f-1)$-ADS in proposed $C^0$, if $n$ is even, the proposed $C^0$ is always near-optimal; otherwise the value of $\rho$ tends to $1$, \emph{i.e.}, the proposed $C^0$ can be asymptotically optimal from near-optimal. Both $C^0$s have $K>2^{x-1}M, (x\geq2)$. The numerical analysis is shown in Tables 2 and 3.
\end{remark}
\begin{eqnarray*}
\begin{tabular}{ccccc}
\multicolumn{5}{c}{Table\ 1:\ For\ the\ Singer\ Difference\ Set} \\
\hline
q & K & M & $K/M$ & $\rho$ \\
\hline
$2^{n}-1$ & $2^n$ & $2^{n-1}-1$ & $\geq 2$ & $1.000$ \\
$2^{n-1}-1$ & $2^n$ & $2^{n-2}-1$ & $\geq 2^2$ & $1.155$ \\
$2^{n-2}-1$ & $2^n$ & $2^{n-3}-1$ & $\geq 2^3$ & $1.512$ \\
$2^{n-3}-1$ & $2^n$ & $2^{n-4}-1$ & $\geq 2^4$ & $2.066$ \\
$2^{n-4}-1$ & $2^n$ & $2^{n-5}-1$ & $\geq 2^5$ & $2.874$ \\
\hline
\end{tabular}
\end{eqnarray*}
\begin{eqnarray*}
\begin{tabular}{ccccc}
\multicolumn{5}{c}{Table\ 2:\ For\ Even $n$\ in\ ADS} \\
\hline
f & K & M & $K/M$ & $\rho$ \\
\hline
$2^{n-2}-1$ & $2^n$ & $2^{n-1}-3$ & $\geq 2$ & $2.000$ \\
$2^{n-3}-1$ & $2^n$ & $2^{n-2}-3$ & $\geq 2^2$ & $1.633$ \\
$2^{n-4}-1$ & $2^n$ & $2^{n-3}-3$ & $\geq 2^3$ & $1.512$ \\
$2^{n-5}-1$ & $2^n$ & $2^{n-4}-3$ & $\geq 2^4$ & $1.461$ \\
$2^{n-6}-1$ & $2^n$ & $2^{n-5}-3$ & $\geq 2^5$ & $1.437$ \\
$2^{n-7}-1$ & $2^n$ & $2^{n-6}-3$ & $\geq 2^6$ & $1.425$ \\
\vdots & \vdots & \vdots & \vdots & \vdots \\
$2^{n-10}-1$ & $2^n$ & $2^{n-9}-3$ & $\geq 2^9$ & $1.416$ \\
\vdots & \vdots & \vdots & \vdots & \vdots \\
$2^{n-20}-1$ & $2^n$ & $2^{n-19}-3$ & $\geq 2^{19}$ & $1.414$ \\
\vdots & \vdots & \vdots & \vdots & \vdots \\
$2^{n-40}-1$ & $2^n$ & $2^{n-39}-3$ & $\geq 2^{39}$ & $1.414$ \\
\hline
\end{tabular}
\end{eqnarray*}
\begin{eqnarray*}
\begin{tabular}{ccccc}
\multicolumn{5}{c}{Table\ 3:\ For\ Odd $n$\ in\ ADS} \\
\hline
f & K & M & $K/M$ & $\rho$ \\
\hline
$2^{n-2}-1$ & $2^n$ & $2^{n-1}-3$ & $\geq 2$ & $1.414$ \\
$2^{n-3}-1$ & $2^n$ & $2^{n-2}-3$ & $\geq 2^2$ & $1.155$ \\
$2^{n-4}-1$ & $2^n$ & $2^{n-3}-3$ & $\geq 2^3$ & $1.069$ \\
$2^{n-5}-1$ & $2^n$ & $2^{n-4}-3$ & $\geq 2^4$ & $1.033$ \\
$2^{n-6}-1$ & $2^n$ & $2^{n-5}-3$ & $\geq 2^5$ & $1.016$ \\
$2^{n-7}-1$ & $2^n$ & $2^{n-6}-3$ & $\geq 2^6$ & $1.008$ \\
\vdots & \vdots & \vdots & \vdots & \vdots \\
$2^{n-10}-1$ & $2^n$ & $2^{n-9}-3$ & $\geq 2^9$ & $1.001$ \\
\vdots & \vdots & \vdots & \vdots & \vdots \\
$2^{n-20}-1$ & $2^n$ & $2^{n-19}-3$ & $\geq 2^{19}$ & $1.000$ \\
\vdots & \vdots & \vdots & \vdots & \vdots \\
$2^{n-40}-1$ & $2^n$ & $2^{n-39}-3$ & $\geq 2^{39}$ & $1.000$ \\
\hline
\end{tabular}
\end{eqnarray*}
Obviously, from Tables 1-3, compared with the proposed QCSS $C^1$ on the signer difference set, the proposed QCSS $C^0$ on the almost difference set has advantages in tightness and supporting more CDMA users.

\section{Summary and concluding remarks}

In this paper, we propose a near-optimal periodic $(2^n, 2^{n-1}-3, 2^n-1)$-QCSS $C^0$ from the Family $\mathcal{A}$ and the $(2^n-4,\ 2^{n-1}-3,\ 2^{n-2}-3,\ 2^{n-2}-2)$-ADS, which has the following periodic tolerance
\begin{eqnarray*}
\delta_{max}&=&(1+2^{n/2}) \sqrt{2^{n}-3},
\end{eqnarray*}
and the asymptotic tightness factor $\rho \in (1,2)$. In this case, the near-optimal QCSS has $K>2M$, hence it can be used to support more CDMA users.

Comparing the proposed QCSS $C^1$ on the singer difference set in \cite{12} with the proposed QCSS $C^0$ on the almost difference set in this paper, with the decrease of the number of rows of each element matrix in proposed QCSSs,
the optimal property of $C^1$ would disappear but $C^0$ can be asymptotically optimal from near-optimal, and more CDMA users can be supported by $C^0$, \emph{i.e.,} $K>2^{x-1}M,\ x\geq 2$.

\end{document}